\documentclass[12pt,a4paper]{article}
\usepackage{epsfig}
\usepackage{amsmath}
\usepackage{amssymb}
\usepackage{verbatim}
\usepackage{bm}
\usepackage{dsfont}
\usepackage[footnotesize]{caption}
\usepackage{subfigure}
\usepackage{cancel}
\usepackage{cite}
\usepackage{bbold}

\begin{document}

\thispagestyle{empty}

\begin{flushleft}
DESY 11-173\\
December 2011
\end{flushleft}

\vskip 1.5cm

\begin{center}
{\Large\bf Predicting $\theta_{13}$ and the Neutrino Mass Scale from Quark Lepton Mass Hierarchies}

\vskip 1cm

W.~Buchm\"uller, V.~Domcke, and K.~Schmitz\\[3mm]
{\it{
Deutsches Elektronen-Synchrotron DESY, 22607 Hamburg, Germany}
}
\end{center}

\vskip 0.75cm

\begin{abstract}
Flavour symmetries of Froggatt-Nielsen type can naturally reconcile
the large quark and charged lepton mass hierarchies and the small quark
mixing angles with the observed small neutrino mass hierarchies and their large
mixing angles. We point out that such a flavour structure, together with the
measured neutrino mass squared differences and mixing angles, strongly constrains
yet undetermined parameters of the neutrino sector. Treating unknown
$\mathcal{O}(1)$ parameters as random variables, we obtain surprisingly
accurate predictions for the smallest mixing angle, 
$\sin^2 2\theta_{13} = 0.07^{+0.11}_{-0.05}$, the smallest neutrino mass,
$m_1 = 2.2^{+1.7}_{-1.4} \times 10^{-3}~\text{eV}$, and one Majorana phase,
$\alpha_{21}/\pi = 1.0^{+0.2}_{-0.2}$.
\end{abstract}

\newpage

\section{Introduction}

It remains a theoretical challenge
to explain the observed pattern of quark and lepton masses and mixings,
in particular the striking differences between the quark sector and the 
neutrino sector.
Promising elements of a theory of flavour are grand unification (GUT) based on 
the groups $\mathrm{SU(5)}$, $\mathrm{SO(10)}$ or $\mathrm{E_6}$,
supersymmetry, the seesaw mechanism and additional
flavour symmetries \cite{Raby:2008gh}. A successful example is the 
Froggatt-Nielsen mechanism~\cite{Froggatt:1978nt} based on spontaneously
broken Abelian symmetries, which parametrizes quark and lepton mass ratios and mixings by powers of a small `hierarchy parameter'~$\eta$.
The resulting structure of mass matrices also arises in compactifications of 
higher-dimensional field and string theories, where the parameter $\eta$
is related to the location of matter fields in the compact dimensions
or to vacuum expectation values of moduli fields 
(cf.~\cite{extradimensions}). 

In this article we consider a Froggatt-Nielsen symmetry which commutes with
the GUT group $\mathrm{SU(5)}$, and which naturally explains the large
$\nu_{\mu}-\nu_{\tau}$ mixing \cite{Froggatt-Nielsen+SU(5)}.
This symmetry implies a particular
hierarchy pattern in the Majorana mass matrix
for the light neutrinos,
\begin{equation}
\label{eq_mnu1}
m_{\nu} \propto
\begin{pmatrix} \eta^{2} & \eta & \eta \\ \eta & 1 & 1 \\ 
\eta & 1 & 1 \end{pmatrix} \ ,
\end{equation}
which can be regarded as a key element for our analysis.
The predicted Dirac and Majorana neutrino mass matrices are also consistent with leptogenesis~\cite{Buchmuller:1998zf}. Despite these successes,
the predictive power of the Froggatt-Nielsen mechanism is rather limited
due to unknown $\mathcal{O}(1)$ coefficients in all entries of the mass matrices.
For example, the considered model \cite{Buchmuller:1998zf} can
accommodate both a small as well as a large `solar' mixing angle  
$\theta_{12}$~\cite{Froggatt-Nielsen+SU(5), Vissani:1998xg}. To get an
idea of the range of possible predictions for a given flavour structure, %one may
it is instructive to 
treat the $\mathcal{O}(1)$ parameters as random 
variables~\cite{random_coeff}.

In the following we shall employ Monte-Carlo techniques to study 
quantitatively the dependence of yet undetermined, but soon testable 
parameters of the neutrino sector on the unknown ${\cal O}(1)$ factors
of the mass matrices. Using the already measured
neutrino masses and mixings as input, we find surprisingly sharp predictions 
which indicate a large value for the smallest mixing angle $\theta_{13}$ 
in accordance with recent results from T2K~\cite{Abe:2011sj}, 
Minos~\cite{Adamson:2011qu} and Double Chooz~\cite{doublechooz}, a value for the lightest neutrino mass
of ${\cal O}(10^{-3})$~eV and one Majorana phase in the mixing
matrix peaked at $\alpha_{21} = \pi$.

\section{Masses and mixings in the lepton sector}

As far as orders of magnitude are concerned, the masses of quarks and charged leptons approximately satisfy the relations
\begin{equation}
 \begin{split}
  &m_t:m_c:m_u \sim 1:\eta^2 : \eta^4 \,, \\
 &m_b : m_s : m_d \sim m_{\tau} : m_{\mu} : m_e \sim 1: \eta : \eta^3 \,,
 \end{split}
\label{eq_masses}
\end{equation}
with $\eta^2 \simeq 1/300$ for masses defined at the GUT scale. This mass 
hierarchy can be reproduced by a simple $\mathrm{U(1)}$ flavour symmetry. 
Grouping the standard model 
leptons and quarks into the $\mathrm{SU(5)}$ multiplets
$\textbf{10} = (q_L, u_R^c, e_R^c)$ and $\textbf{5}^* = (d_R^c, l_L)$, 
the Yukawa interactions take the form
\begin{equation}
\label{eq_L}
 {\cal L}_Y = h_{ij}^{(u)} \textbf{10}_i \textbf{10}_j H_u 
+ h_{ij}^{(e)} \textbf{5}^*_i \textbf{10}_j H_d 
+ h_{ij}^{(\nu)} \textbf{5}_i^* \textbf{1}_j H_u 
+ \frac{1}{2} \, h_{i}^{(n)} \textbf{1}_i \textbf{1}_i S + \mathrm{c.c.}\ ,
\end{equation}
where $\textbf{1} = \nu_{R}^c$ denote the charge conjugates of right-handed neutrinos and 
$i,j = 1\ldots 3$ 
are flavour indices.
Note that the Yukawa matrix $h^{(n)}$ for the right-handed neutrinos
can always be chosen to be real and diagonal.
$H_u$, $H_d$ and $S$ are the Higgs fields for
electroweak and ${B-L}$ symmetry breaking, i.e., their vacuum 
expectation values
generate the Dirac masses of quarks and leptons and the Majorana masses for 
the right-handed neutrinos, respectively. 
In this setup, the Yukawa couplings are determined up to complex ${\cal O}(1)$ factors 
by assigning $\mathrm{U(1)}$ charges to the fermion and Higgs fields in Eq.~\eqref{eq_L},
\begin{equation}
\label{eq_h}
 h_{ij} \sim \eta^{Q_i + Q_j} \ .
\end{equation}
With the charge assignment given in Tab.~\ref{tab_fn-charges} the mass 
relations in Eq.~\eqref{eq_masses} are reproduced. Additionally, perturbativity of the Yukawa couplings and constraints on $\tan \beta  = \langle H_u \rangle / \langle H_d \rangle$ require $0 \leq a \leq 1$.
\begin{table}
\begin{center}
\begin{tabular}{c|cccccccccccc}
$\psi_i$ & $\textbf{10}_3$ & $\textbf{10}_2$ & $\textbf{10}_1$ & 
$\textbf{5}^*_3$ & $\textbf{5}^*_2$ & $\textbf{5}^*_1$ & $\textbf{1}_3$ & 
$\textbf{1}_2$ & $\textbf{1}_1$ & $H_u$ & $H_d$ & $S$  \\
\hline
$Q_i$ & 0& 1 & 2 & $a$ & $a$ & $a+1$ & $b$ & $c$ & $d$ & 0 & 0 & 0
\end{tabular}\end{center}
\caption{Froggatt-Nielsen charge assignments. From Ref. \cite{Buchmuller:1998zf}.}
\label{tab_fn-charges}
\end{table}
\medskip

\newpage \noindent \textit{Masses} \\
\noindent From Eq.~\eqref{eq_L} and Tab.~\ref{tab_fn-charges} one obtains for the Dirac 
neutrino mass matrix $m_D$ and the Majorana mass matrix of the right-handed 
neutrinos $M$,
\begin{equation}\
 \frac{m_D}{v_{EW}\sin\beta} = h^{(\nu)}_{ij} \sim \eta^a 	
\begin{pmatrix} \eta^{d+1} & \eta^{c+1} & \eta^{b+1} \\ 
\eta^{d}& \eta^{c} & \eta^{b} \\ \eta^{d} & \eta^{c} & \eta^{b}\end{pmatrix}\ ,
\quad \frac{M}{v_{B-L}} = h_{ij}^{(n)} \sim  
\begin{pmatrix} \eta^{2d} & 0 & 0 \\ 0 & \eta^{2c} & 0 \\ 
0 & 0  & \eta^{2b}  \end{pmatrix} \ ,
\end{equation}
with the electroweak and ${B-L}$ symmetry breaking vacuum expectation
values $v_{EW}=\sqrt{\langle H_u \rangle^2 + \langle H_d \rangle^2}$ 
and $v_{B-L} = \langle S \rangle$,
respectively.
In the seesaw formula 
\begin{equation}
\label{eq_seesaw}
 m_{\nu} = - m_D \frac{1}{M} m_D^T\ ,
\end{equation}
the dependence on the right-handed neutrino charges drops out, and one finds
for the light neutrino mass matrix,
\begin{equation}
\label{eq_mnu}
 m_{\nu} \sim 
\frac{v_{EW}^2\sin^2\beta}{v_{B-L}} \ \eta^{2a} \  
\begin{pmatrix} \eta^{2} & \eta & \eta \\ \eta & 1 & 1 \\ 
\eta & 1 & 1 \end{pmatrix} \ .
\end{equation}

The charged lepton mass matrix is given by
\begin{equation}
 \frac{m_e}{v_{EW}\cos\beta} = h^{(e)}_{ij} \sim \eta^a 	
\begin{pmatrix} \eta^{3} & \eta^{2 } & \eta \\ 
\eta^{2}& \eta & 1 \\ \eta^{2} & \eta & 1\end{pmatrix}\ .
\end{equation}
Note that the second and third row of the matrix $m_e$ have the same hierarchy pattern. This is a consequence of the same flavour charge for the second and third generation of leptons, which is the origin of the large neutrino mixing. Hence, diagonalizing $m_e$ can a priori give a sizable contribution to the mixing in the lepton sector. \medskip

\noindent\textit{Mixing} \\ \noindent
The lepton mass matrices are diagonalized by bi-unitary and unitary 
transformations, respectively,
\begin{equation}
V_L^{T} m_e V_R = m_e^{\mathrm{diag}}\ , \quad
U^T m_{\nu} U = m_{\nu}^{\mathrm{diag}} \ ,
\end{equation}
with $V_L^{\dagger}V_L = V_R^{\dagger}V_R = U^{\dagger}U = \mathbb{1}$.
From $V_L$ and $U$ one obtains the leptonic mixing matrix
$U_{\mathrm{PMNS}} = V_L^{\dagger}U$,
which is parametrized as \cite{Nakamura:2010zzi}
\begin{align}\label{pmns}
U_{\mathrm{PMNS}} =
\begin{pmatrix} c_{12}c_{13} & s_{12}c_{13}e^{i\frac{\alpha_{21}}{2}}
 & s_{13}e^{i(\frac{\alpha_{31}}{2}-\delta)} \\ 
-s_{12}c_{23}-c_{12}s_{23}s_{13}e^{i\delta} & 
\left(c_{12}c_{23}
-s_{12}s_{23}s_{13}e^{i\delta}\right)e^{i\frac{\alpha_{21}}{2}} &
s_{23}c_{13}e^{i\frac{\alpha_{31}}{2}} \\
s_{12}s_{23}-c_{12}c_{23}s_{13}e^{i\delta} &
\left(-c_{12}s_{23}
-s_{12}c_{23}s_{13}e^{i\delta}\right)e^{i\frac{\alpha_{21}}{2}} &
 c_{23}c_{13}e^{i\frac{\alpha_{31}}{2}}\end{pmatrix}\ ,
\end{align}
with $c_{ij} = \cos \theta_{ij}$ and $s_{ij} = \sin \theta_{ij}$. Since the light neutrinos are Majorana fermions, all three phases are physical. 

In the following we study the impact of the unspecified ${\cal O}(1)$ factors
in the lepton mass matrices on the various parameters of the neutrino sector
by using a Monte Carlo method, taking present knowledge on neutrino masses 
and mixings into account.
Naively, one might expect large uncertainties in the predictions for the observables of the neutrino sector obtained in this setup. For instance, the neutrino mass matrix is calculated by multiplying three matrices, in which each entry comes with an unspecified ${\cal O}(1)$ factor, cf.\ Eq.~\eqref{eq_seesaw}. However, carrying out the analysis described below and calculating the $68\%$ confidence intervals, we find that in many cases our results are sharply peaked, yielding a higher precision than only an order-of-magnitude estimate.

\section{Random variables}
\noindent \textit{Monte-Carlo study}

\noindent The unknown ${\cal O}(1)$ coefficients of the Yukawa matrices $h^{(e)}$, 
$h^{(\nu)}$ and $h^{(n)}$ are constrained by the experimental data on neutrino 
masses and mixings, with the $3 \sigma$ confidence ranges given by~\cite{Nakamura:2010zzi}:
\begin{equation}
 \label{eq_exp}
\begin{split}
& 2.07 \times 10^{-3} \, \text{eV}^2 \leq |\Delta m^2_{\text{atm}}| \leq 2.75 \times 10^{-3} \, \text{eV}^2 \,, \\
& 7.05 \times 10^{-5} \, \text{eV}^2 \leq  \Delta m^2_{\text{sol}} \leq 8.34 \times 10^{-5} \, \text{eV}^2 \,, \\
& 0.75 \leq \sin^2(2 \theta_{12})  \leq 0.93 \,, \\
& 0.88 \leq  \sin^2(2 \theta_{23}) \leq 1 \,.
\end{split}
\end{equation}
In the following we explicitly do not use the current bound on the smallest mixing angle ($\theta_{13} < 0.21$ at $3 \sigma$ \cite{Nakamura:2010zzi}). This allows us to demonstrate that nearly all values we obtain for $\theta_{13}$ automatically obey the experimental bound, cf.~Fig.~\ref{fig_mixing}. 

In a numerical Monte-Carlo study we generate random numbers to model the 
39 real parameters of the three mass matrices.\footnote{Nine complex ${\cal O}(1)$ factors in each $h^{(\nu)}$ and $h^{(e)}$, as well as three real ${\cal O}(1)$ factors in $h^{(n)}$. Note that here we are treating the low energy Yukawa couplings as random variables, which are related to the couplings at higher energy scales via renormalization group equations. However, we expect that the effect of this renormalization group running can essentially be absorbed into a redefinition of the effective scale $\bar{v}_{B-L}$, hence leaving the results presented in the following unchanged.} The absolute values are taken 
to be uniformly distributed in $ [10^{-1/2}, 10^{1/2}]$ on a logarithmic 
scale. The phases in $h^{(e)}$ and $h^{(\nu)}$ are chosen to be uniformly 
distributed in $ [0, 2 \pi)$. In the following, we shall refer to those sets 
of coefficients which are consistent with the 
experimental constraints in Eq.~\eqref{eq_exp} as hits.

In a preliminary run, we consider the neutrino mixing matrix $U$, 
with the effective scale $\bar{v}_{B-L} \equiv \eta^{-2a} v_{B-L}/\sin^2\beta$ 
treated as random variable in the interval 
$[10^{-1/2}, 10^{1/2}]\times10^{15}$~GeV. We find that the percentage of hits 
strongly peaks at $\bar{v}_{B-L} \simeq 1 \times 10^{15}$~GeV. This is 
interesting for two reasons. Firstly, it implies that given $0 \leq a \leq 1$, 
the high seesaw scale lies in the range 
$3\times 10^{12}~\text{GeV}\lesssim v_{B-L}/\sin^2\beta\lesssim 
1\times 10^{15}~\text{GeV}$.
Note that the upper part of this mass range is close the GUT scale, which is 
important for recent work on the connection of leptogenesis, gravitino dark
matter and hybrid inflation 
\cite{cosmo}.
Secondly, this result allows us to fix the parameter $\bar{v}_{B-L}$ in the 
following computations without introducing a significant bias.

In the main run, for fixed $\bar{v}_{B-L}$, we include the mixing matrix $V_L$ of 
the charged leptons to compute the full PMNS matrix. We require the mass ratios 
of the charged leptons to fulfill the experimental constraints up to an accuracy 
of $5\%$ and allow for ${1\leq\tan\beta \leq 60}$ to achieve the correct 
normalization of the charged lepton mass spectrum. Finally, imposing the 
$3\sigma$ constraints on the two large mixing angles of the full PMNS matrix, we find 
parameter sets of ${\cal O}(1)$ factors which yield mass matrices
fulfilling the constraints in Eq.~\eqref{eq_exp}. Our final results are based on roughly
20\,000 such hits. For each hit we calculate the observables in the neutrino
sector as well as parameters relevant for leptogenesis. The resulting 
distributions are discussed below.\medskip

\noindent \textit{Statistical analysis}

\noindent In our theoretical setup the relative frequency with which we encounter a 
certain value for an observable might indicate the probability that this value 
is actually realized within the large class of flavour models under study.
In the following we shall therefore treat the distributions for the various 
observables as probability densities for continuous random variables.
That is, our predictions for the respective observables represent best-guess 
estimates according to a probabilistic interpretation of the relative 
frequencies.

For each observable we would like to deduce measures
for its central tendency and statistical dispersion from
the respective probability distribution.
Unfortunately, it is infeasible to fit all obtained distributions
with one common template distribution.
Such a procedure would lack a clear statistical justification, and it also
appears impractical as the distributions that we obtain differ substantially 
in their shapes.
We therefore choose a different approach.
We consider the median of a distribution as its centre
and we use the $68\ \%$ `confidence' interval around it
as a measure for its spread.
Of course, this range of the confidence interval is reminiscent of
the $1\sigma$ range of a normal distribution.

More precisely, for an observable
$x$ with probability density $f$ we will
summarize its central tendency and variability in the following
form \cite{Cowan:1998ji},
\begin{align}
x = \hat{x}_{\Delta_-}^{\Delta_+}
\,,\quad
\Delta_\pm = x_\pm - \hat{x}
\,.
\end{align}
Here, $x_-$ and $x_+$ denote the $16\,\%$- and $84\,\%$-quantiles with respect to the density function
$f$.
The central value $\hat{x}$ is the median of $f$ and thus
corresponds to its $50\,\%$-quantile.
All three values of $x$ can be calculated from
the quantile function $Q$,
\begin{align}
Q(p) = \textrm{inf} \left\{x \in
\left[x_\textrm{min},x_\textrm{\text{max}}\right] : p \leq F(x)\right\}
\,,\quad
F(x) = \int_{x_\textrm{min}}^x dt \: f(t)
\,,
\end{align}
where $F$ stands for the cumulative distribution function of $x$.
We then have:
\begin{align}
x_- = Q(0.16)\,,\quad
\hat{x} = Q(0.50)\,,\quad
x_+ = Q(0.84)\,.
\end{align}
Intuitively, the intervals from $x_{\textrm{min}}$ to $x_-$, $\hat{x}$, and $x_+$ respectively correspond to
the $x$ ranges into which $16\,\%$, $50\,\%$ or $84\,\%$
of all hits fall.
This is also illustrated in the histogram
for $\sin^2 2\theta_{13}$ in Fig.~1.
Moreover, we have included vertical lines into each plot
to indicate the respective positions of $x_-$, $\hat{x}$, and $x_+$.

In our case the median is a particularly useful measure
of location.
First of all, it is resistant against outliers and hence an appropriate statistic for such skewed distributions as we observe them.
But more importantly, the average absolute deviation from
the median is minimal in comparison to any other reference point.
The median is thus the best guess for the outcome of a
measurement if one is interested in being as close as possible
to the actual result, irrespective of the sign of the error.
On the technical side the definition of the median fits nicely
together with our method of assessing statistical dispersion.
The $68\,\%$ confidence interval as introduced above
is just constructed in such a way that equal numbers of hits
lie in the intervals from $x_-$ to $\hat{x}$ and from $\hat{x}$
to $x_+$, respectively.
In this sense, our confidence interval represents a symmetric
error with respect to the median.

As a test of the robustness of our results, we checked the dependence of our distributions on the precise choice of the experimental error intervals. The results presented here proved insensitive to these variations. For definiteness, we therefore stick to the $3 \sigma$ intervals. We also checked the effect of taking the random ${\cal{O}}(1)$ factors to be distributed uniformly on a linear instead of a logarithmic scale. Again, the results proved to be robust.

\section{Observables and results}

\noindent \textit{Mass hierarchy}

\noindent An important open question which could help unravel the flavour structure of 
the neutrino sector is the mass hierarchy. Since the sign of $\Delta m^2_{\text{atm}}$ 
is not yet known, we cannot differentiate with current experimental data between
a normal hierarchy with one heavy and two light neutrino mass eigenstates and 
an inverted hierarchy, which has two heavy and one light neutrino mass eigenstate.
Measuring the Mikheyev-Smirnov-Wolfenstein (MSW) effect of the earth could 
resolve this ambiguity.

With the procedure described above, all hits match the structure of the normal 
hierarchy and there are no examples with inverted hierarchy. It is however 
notable that imposing the structure of the neutrino mass matrix given by 
Eq.~\eqref{eq_mnu} alone does not exclude the inverted mass hierarchy. Only 
additionally imposing the measured bounds on the mixing angles rejects this 
possibility.\medskip

\begin{figure}
\subfigure{
 \includegraphics[width=0.48\textwidth]{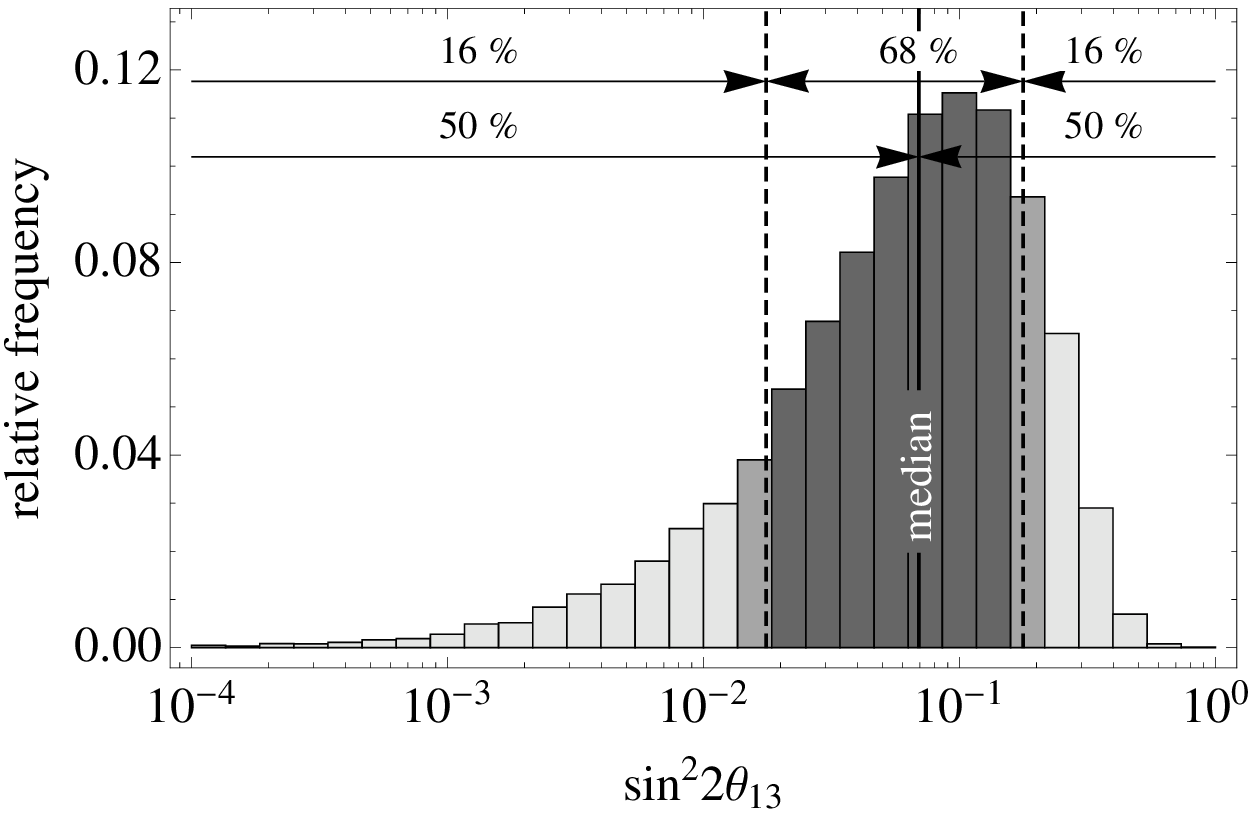}} 
\subfigure{
 \includegraphics[width=0.48\textwidth]{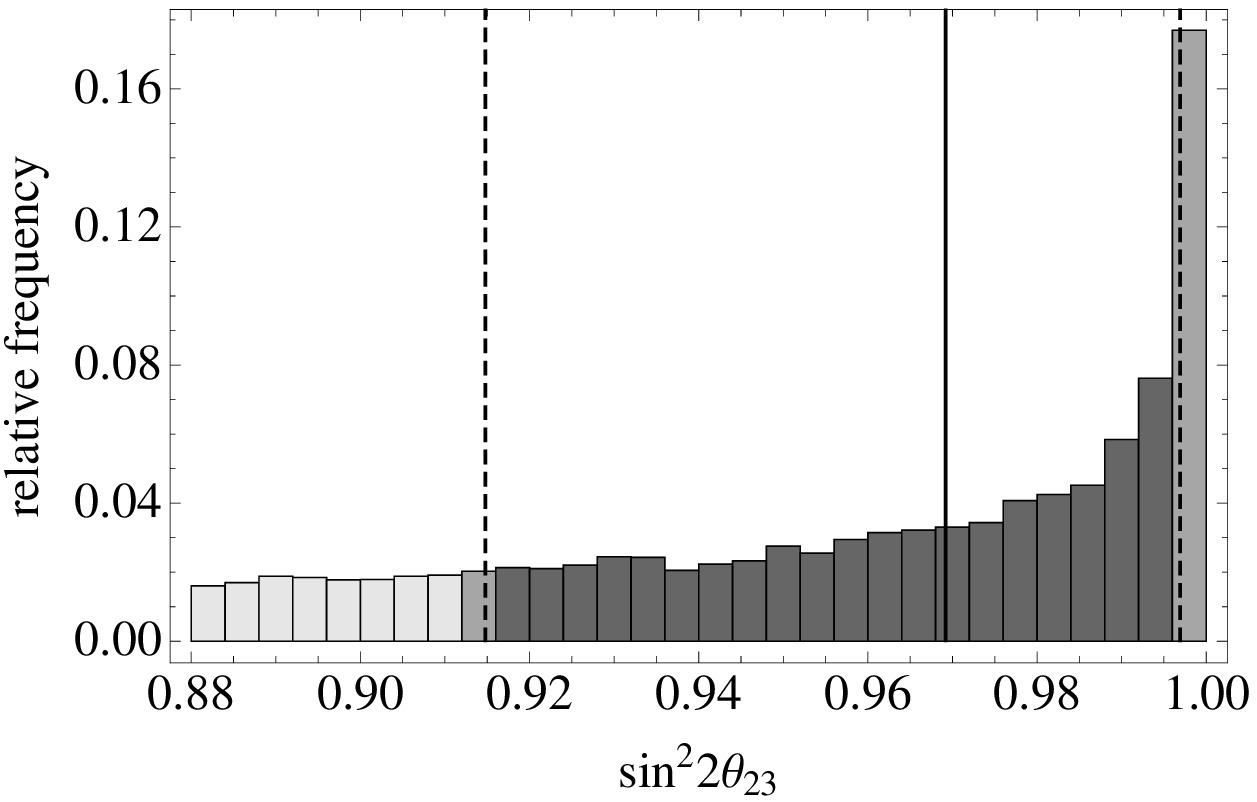}}
\caption{Neutrino mixing angles $\theta_{13}$ and $\theta_{23}$. The vertical lines denote the position of the median (solid line) and the boundaries of the $68 \%$ confidence region (dashed lines) of the respective distribution.}
\label{fig_mixing}
\end{figure}

\noindent\textit{Mixing angles}

\noindent The mixing in the lepton sector is described by the matrix $U_{\mathrm{PMNS}}$
given in Eq.~\eqref{pmns}. Of the three angles, two are only bounded from one 
side by experiment: for the largest mixing angle $\theta_{23}$ there exists  
a lower bound, whereas the smallest mixing angle $\theta_{13}$ is so far only 
bounded from above. Recent results from T2K~\cite{Abe:2011sj}, Minos~\cite{Adamson:2011qu} and the preliminary result of Double Chooz~\cite{doublechooz}
point to a value of 
$\theta_{13}$ just below the current experimental bound. The respective best 
fit points, assuming a normal hierarchy, are  
$\sin^2 2\theta_{13} = 0.11$ (T2K),
$2 \sin^2 \theta_{23} \sin^2 2 \theta_{13} = 0.041$ (MINOS) and $\sin^2 2 \theta_{13} = 0.085$ (Double Chooz). The 90$\%$ and 68$\%$ confidence regions respectively
read
\begin{align}
 &0.03 < \sin^2 2\theta_{13} < 0.28   &&\text{T2K, 90 $\%$ CL}, \, \delta_{CP} = 0, \nonumber \\
&2 \sin^2 \theta_{23} \sin^2 2 \theta_{13} < 0.12  
\qquad &&\text{MINOS, 90 $\%$ CL}, \, \delta_{CP} = 0,  \\
& 0.01 < \sin^2 2\theta_{13} < 0.16  &&\text{Double Chooz, 68 $\%$ CL}. \nonumber
\end{align}

With the procedure described above, we find sharp predictions for the smallest 
and the largest mixing angle within the current experimental bounds, 
\begin{equation}
 \sin^2 2\theta_{13} = 0.07^{+0.11}_{-0.05}  \ , 
\qquad  \sin^2 2\theta_{23} = 0.97^{+0.03}_{-0.05}  \ ;
\end{equation}
the corresponding distributions are shown in Fig.~\ref{fig_mixing}.
These results are quite remarkable: the atmospheric mixing angle points
to maximal mixing, while the rather large value for $\theta_{13}$ is consistent
with the recent T2K, Minos and Double Chooz results.

In our Monte-Carlo study we observe that the dominant contribution to the strong mixing in the 
lepton sector is primarily due to the neutrino mass matrix $m_{\nu}$.
The numerical results are not much affected by including the charged lepton 
mixing matrix $V_L$. The PMNS matrix is thus approximately given by the matrix 
$U$ which diagonalizes the light neutrino mass matrix $m_{\nu}$. \medskip

\begin{figure}[t]
\subfigure{
 \includegraphics[width=0.48\textwidth]{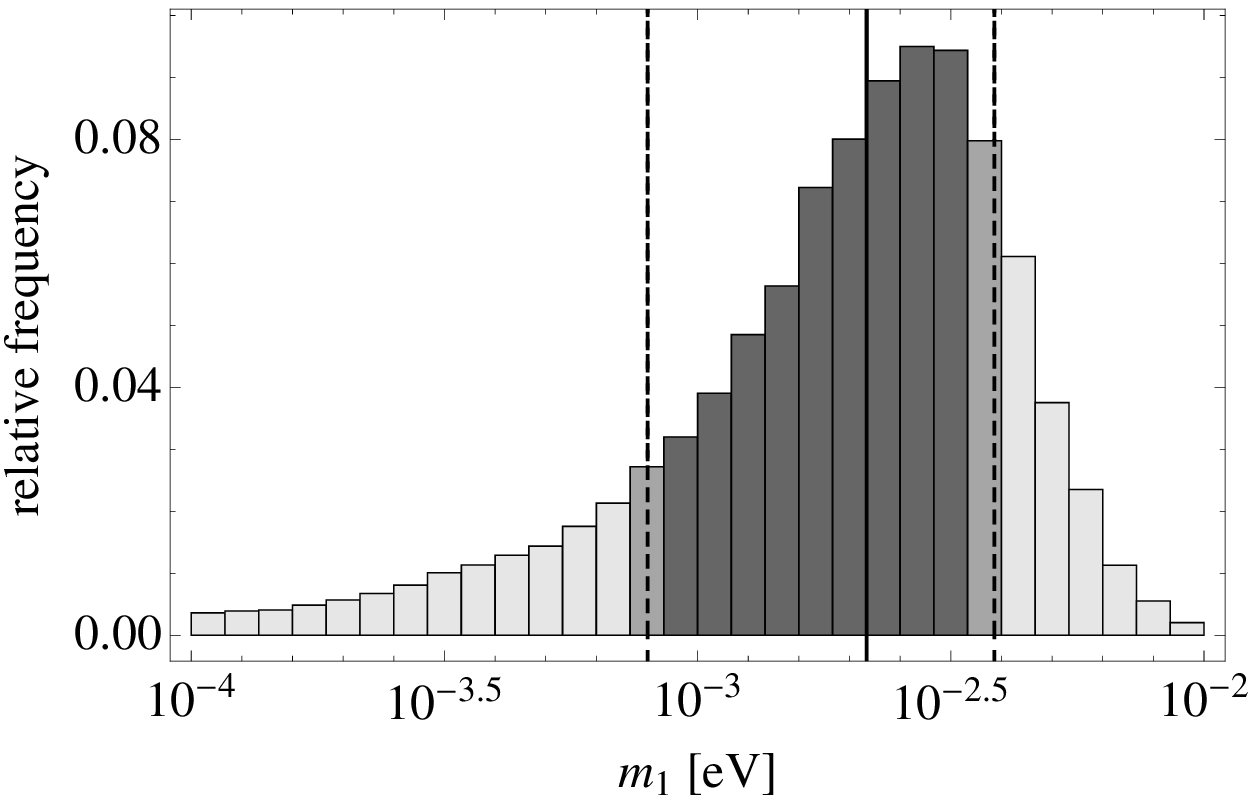}}
\subfigure{
 \includegraphics[width=0.48\textwidth]{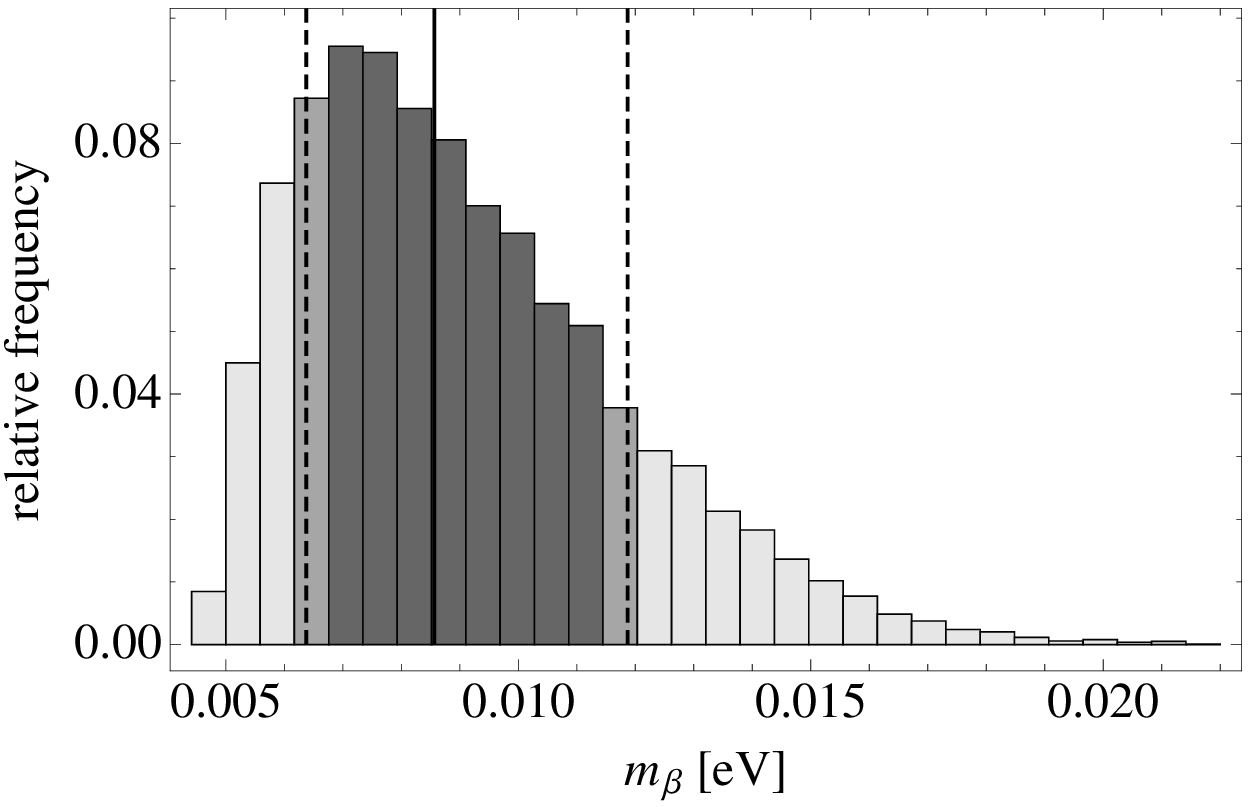}}
\caption{Lightest neutrino mass $m_1$ and effective neutrino mass in tritium decay $m_\beta$. Vertical lines and shadings as in Fig.~\ref{fig_mixing}.}
\label{fig_m1mtritium}
\end{figure}

\newpage
\noindent \textit{Absolute mass scale}

\noindent The absolute neutrino mass scale is a crucial ingredient for the study of neutrinoless double-beta decay and leptogenesis. Although inaccessible in neutrino oscillation experiments, different experimental setups have succeeded in constraining this mass scale. Cosmological observations of the fluctuations in the cosmic microwave background, of the density fluctuations in the galaxy distribution and of the Lyman-$\alpha$ forest yield a constraint for the sum of the light neutrino masses, weighted by the number of spin degrees of freedom per Majorana neutrino, $g_{\nu}=2$, \cite{Nakamura:2010zzi}
\begin{equation}
 m_{\text{tot}} = \sum_{\nu} \frac{g_{\nu}}{2} m_{\nu} \lesssim 0.5 \, \text{eV} \,.
\end{equation}
\begin{figure}[t]
\subfigure{
 \includegraphics[width=0.48\textwidth]{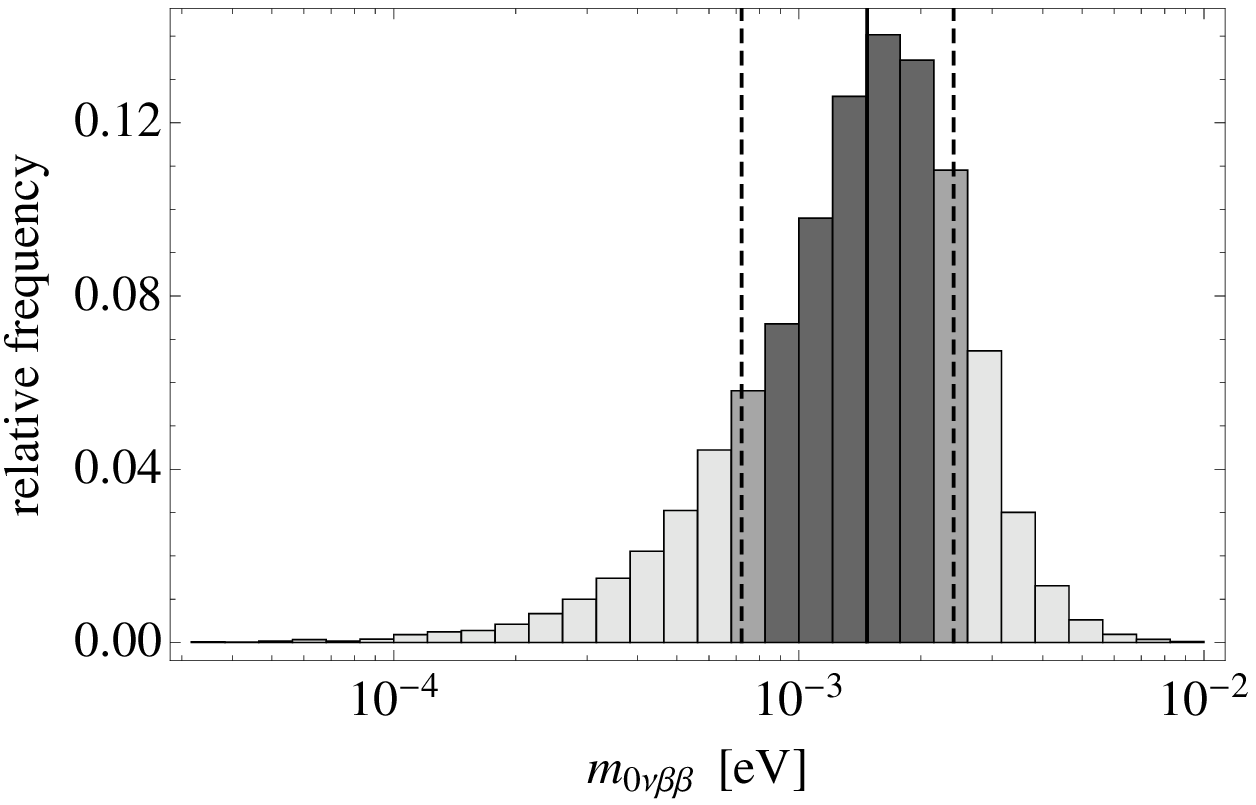}} 
\subfigure{
 \includegraphics[width=0.48\textwidth]{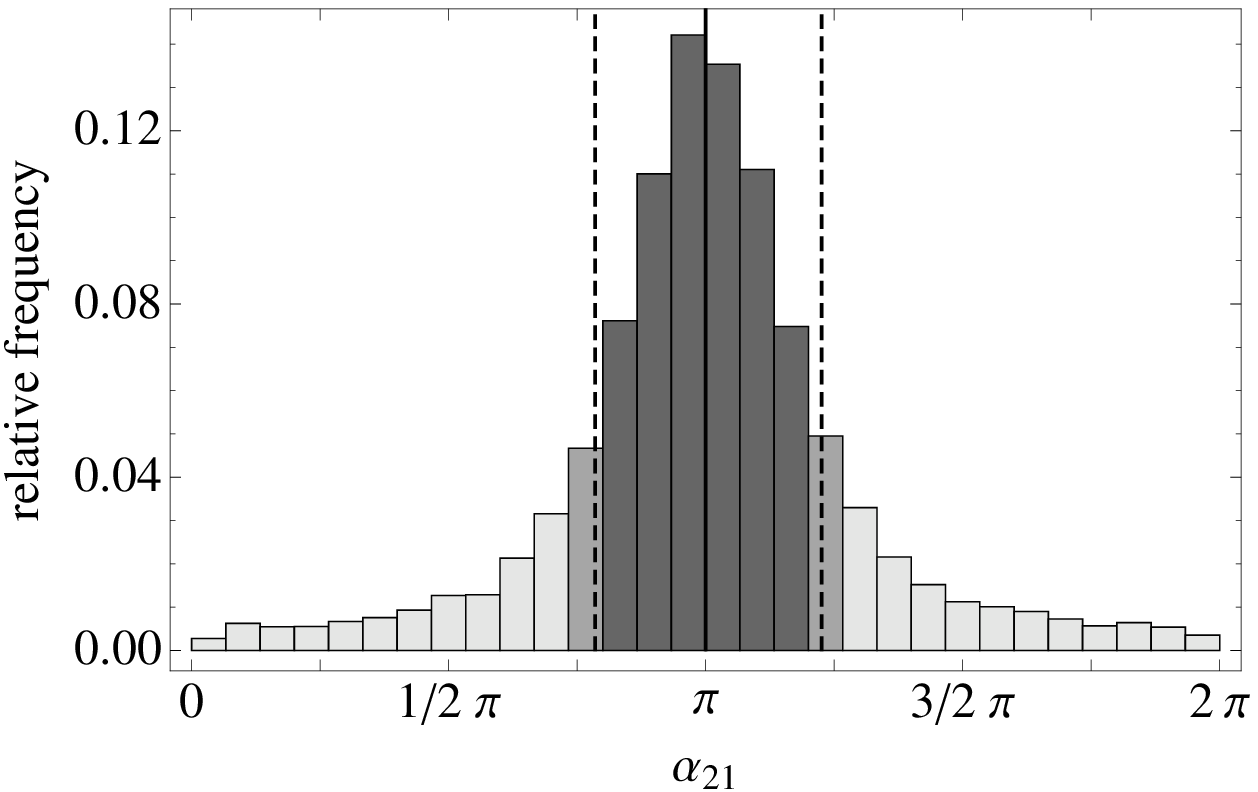}}
\caption{Effective mass in neutrinoless double-beta decay $m_{0\nu\beta\beta}$ and Majorana phase $\alpha_{21}$. Vertical lines and shadings as in Fig.~\ref{fig_mixing}.}
\label{fig_masses}
\end{figure} 
The Planck satellite is expected to be sensitive to values of $m_{\text{tot}}$ as low as roughly $0.1$~eV~\cite{:2006uk}. A further constraint arises from measuring the $\beta$-spectrum in tritium decay experiments. The current bound~\cite{Nakamura:2010zzi} is
\begin{equation}
  m_{\beta}^2  = \sum_{i} |(U_{PMNS})_{ei}|^2 m_{i}^2 < 4~\text{eV}^2 \,.
\end{equation}
By comparison, the KATRIN experiment, which will start taking data soon, aims at reaching a sensitivity of $0.04 \,\text{eV}^2$~\cite{Beck:2010zzb}. Finally, the neutrino mass scale can also be probed by neutrinoless double-beta decay. The relevant effective mass is
\begin{equation}
 m_{0\nu\beta\beta} = |\sum_i (U_{PMNS})^2_{ei} m_i | \,.
\label{eq_m0nubb}
\end{equation}
Here, Ref.~\cite{KlapdorKleingrothaus:2001ke} claims a value of $0.11 - 0.56$~eV. Dedicated experiments, such as GERDA~\cite{Meierhofer:2011zz} with a design sensitivity of $0.09 - 0.20$~eV, are on the way. Note that $m_{0\nu\beta\beta}$ does
 not only depend on the absolute neutrino mass scale and the mixing angles, but also on the phases $(\alpha_{31} - 2 \delta)$ and $\alpha_{21}$ in the PMNS matrix.

We find sharp predictions for the neutrino mass parameters discussed above. The lightest neutrino, $\nu_1$, is found to be quite light, cf.\ Fig.~\ref{fig_m1mtritium},
\begin{equation}
 m_1 = 2.2^{+1.7}_{-1.4} \times 10^{-3} \, \text{eV}\,,
\end{equation}
hence favouring a relatively low neutrino mass scale beyond the reach of current and upcoming experiments. More precisely, we find for the neutrino mass parameters discussed above:
\begin{equation}
m_{\text{tot}} = 6.0^{+0.3}_{-0.3} \times 10^{-2} \, \text{eV} , \quad   m_{\beta} = 8.6^{+3.3}_{-2.2} \times 10^{-3}  \, \text{eV} , \quad  m_{0\nu\beta\beta} = 1.5^{+0.9}_{-0.8} \times 10^{-3}  \, \text{eV} .
\end{equation}

\noindent \textit{CP-violating phases}

\noindent The small value of the mass parameter measured in neutrinoless double-beta decay, $m_{0\nu\beta\beta}$, is due to the relative minus sign between the $m_1$ and $m_2$ terms in Eq.~\eqref{eq_m0nubb}, caused by a strong peak of the value for the Majorana phase $\alpha_{21}$ at $\pi$,
\begin{equation}
 \frac{\alpha_{21}}{\pi} = 1.0^{+0.2}_{-0.2}  \,.
\end{equation}
This is depicted in Fig.~\ref{fig_masses}. An analytic analysis of how this phenomena arises from the structure of the neutrino mass matrix, cf.\ Eq.~\eqref{eq_mnu}, is presented in Appendix~\ref{app_alpha1}. For the other Majorana phase $\alpha_{31}$ and the Dirac phase $\delta$ we find no such distinct behaviour but approximately flat distributions. \medskip

\begin{figure}[t]
\subfigure{
 \includegraphics[width=0.48\textwidth]{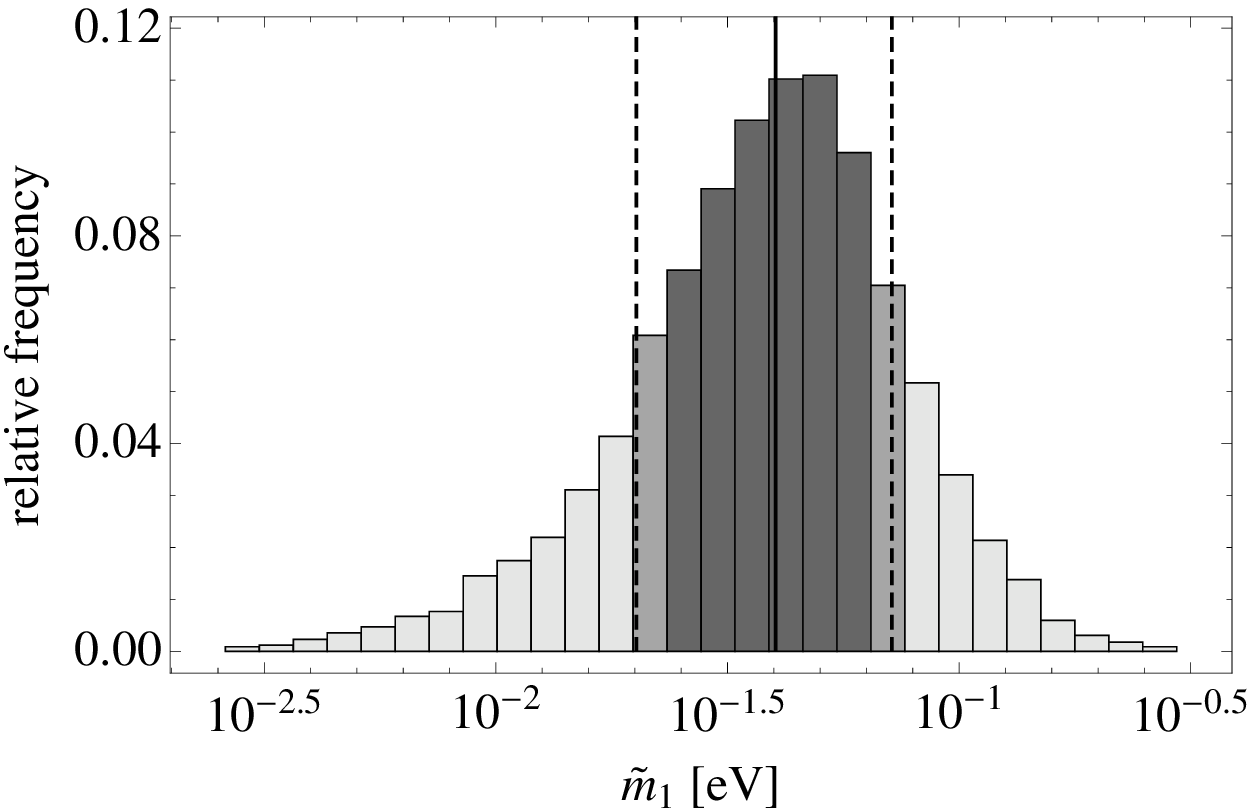}} %\hfill \hspace{0.3cm}
\subfigure{
 \includegraphics[width=0.48\textwidth]{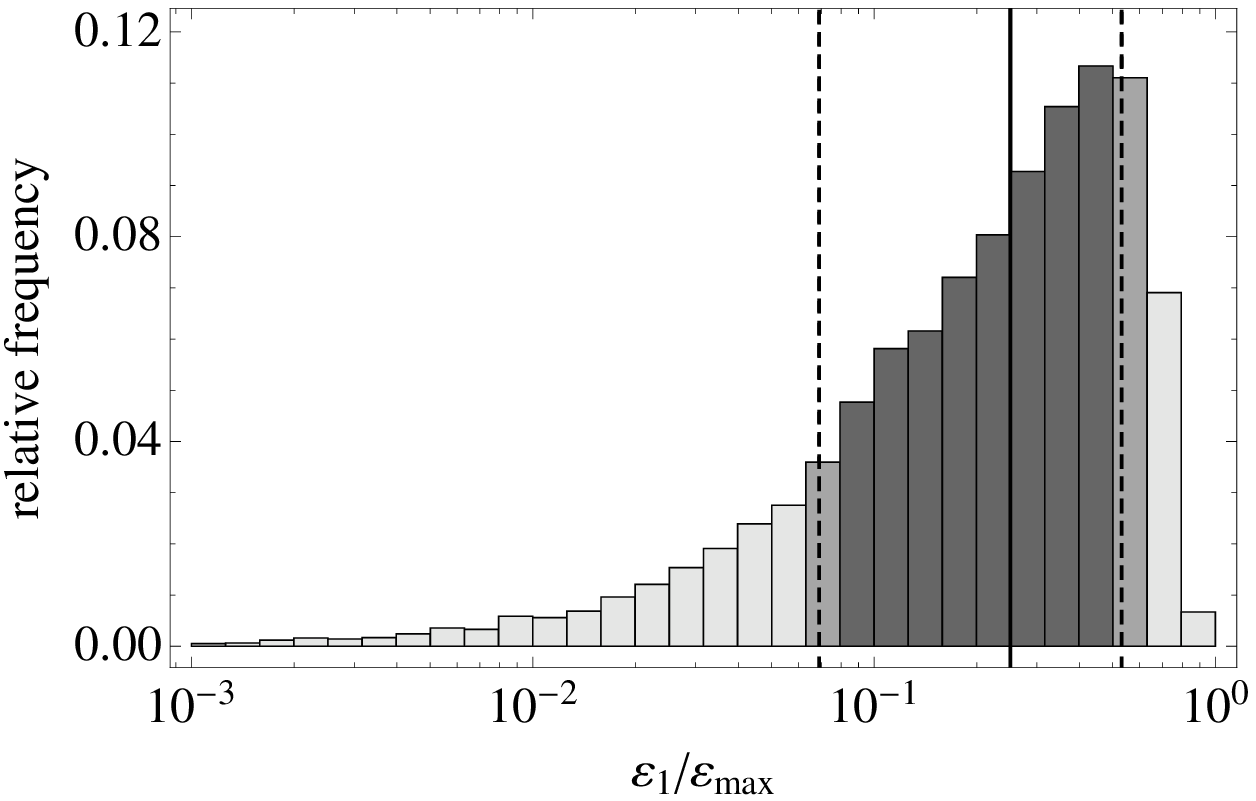}}
\caption{Effective neutrino mass of the first generation $\widetilde{m}_1$ and $CP$ violation parameter $\varepsilon_1$. Vertical lines and shadings as in Fig.~\ref{fig_mixing}.}
\label{fig_CP}
\end{figure} 

\newpage
\noindent \textit{Leptogenesis parameters}

\noindent Finally, leptogenesis~\cite{Fukugita:1986hr} links the low energy neutrino physics to the high energy physics of the early universe. The parameters that capture this connection are the effective neutrino mass of the first generation $\widetilde{m}_1$ and the $CP$ violation parameter $\varepsilon_1$~\cite{leptogenesis},
\begin{equation}
\widetilde{m}_1 = \frac{(m^{\dagger}_D m_D)_{11}}{M_1}\,, \qquad
\varepsilon_1  =  - \sum_{j=2,3}\frac{\text{Im}  \left[ (h^{(\nu) \, \dagger} h^{(\nu)})_{1j} \right]^2 }{8 \pi (h^{(\nu) \, \dagger} h^{(\nu)})_{11}} F\left(\frac{M^2_j}{M^2_{1}} \right)\,,
\end{equation}
with $F(x) = \sqrt{x} \left(\text{ln} \frac{1+x}{x} + \frac{2}{x-1} \right)$ and $M_{j}$ denoting the masses of the heavy neutrinos. Here, $\widetilde{m}_1$ determines the coupling strength of the lightest of the heavy neutrinos to the thermal bath and thus controls the significance of wash-out effects. It is bounded from below by the lightest neutrino mass $m_1$. The absolute value of the $CP$ violation parameter $\varepsilon_1$ is bounded from above by~\cite{epsilon}
\begin{equation}
 \varepsilon_{\text{max}} = \frac{3}{8 \pi}\frac{|\Delta m_{\text{atm}}^2|^{1/2} \, M_1 }{ v^2_{EW} \sin^2 \beta}  \simeq 2.1 \times 10^{-6} \, \left( \frac{1}{\sin^2 \beta} \right) \left(\frac{M_1}{10^{10} \, \text{GeV}} \right).
\end{equation}
With the procedure described above, we find
\begin{equation}
  \widetilde{m}_1 = 4.0^{+3.1}_{-2.0} \times 10^{-2} \, \text{eV} \,, \qquad  \frac{\varepsilon_1}{\varepsilon_{\text{max}}} = 0.25^{+0.28}_{-0.18}\,,
\end{equation}
and hence a clear preference for the strong wash-out regime~\cite{leptogenesis}. Notice that there typically is a hierarchy between $\widetilde{m}_1$ and $m_1$ of about one order of magnitude. The relative frequency of the $CP$ violation parameter $\varepsilon_1$ peaks close to the upper bound $\varepsilon_{\text{max}}$, with the majority of the hits lying within one order of magnitude or less below $\varepsilon_{\text{max}}$, cf.\ Fig.~\ref{fig_CP}. This justifies the use of $\varepsilon_{\text{max}}$ when estimating the produced lepton asymmetry in leptogenesis. Here, in the discussion of $\varepsilon_1$, we assumed hierarchical heavy neutrinos, $M_{2,3} \gg M_1$. \medskip

\noindent \textit{Theoretical versus experimental input}\\
\noindent The results of this section are obtained by combining two conceptually different inputs, on the one hand the hierarchy structure of the neutrino mass matrix $m_{\nu}$ given by Eq.~\eqref{eq_mnu1} and on the other hand the experimentally measured constraints listed in Eq.~\eqref{eq_exp}. In general, the distributions presented above really arise from the interplay between both of these ingredients. For example, the hierarchy structure alone does not favour a large solar mixing angle $\theta_{12}$ and the ratio $\Delta m^2_{\text{sol}} / \Delta m^2_{\text{atm}}$ tends to be too large (cf.~\cite{masina,winter}). This discrepancy is eased by generating the random coefficients in Eq.~\eqref{eq_mnu1} via the seesaw mechanism. Imposing the experimental constraints finally singles out the subset of parameter sets used for the distributions presented above. As another example, consider the smallest mixing angle $\theta_{13}$ and the lightest neutrino mass eigenstate $m_1$. In these cases, the hierarchy structure of the neutrino mass matrix automatically implies small values, similar to those shown in the distributions above. However, the exact distributions including the precise position of the peaks only arise after implementing the experimental constraints. A notable exception to this scheme is the Majorana phase $\alpha_{21}$. Here the peak at $\alpha_{21} = \pi$ is a result of the hierarchy structure of the neutrino matrix $m_{\nu}$ alone, as demonstrated in Appendix~\ref{app_alpha1}.

\section{Discussion and outlook}
In summary, we find that starting from a flavour symmetry which accounts for
the measured quark and lepton mass hierarchies and large neutrino mixing, the present knowledge of
neutrino parameters strongly constrains the yet unknown observables, in particular the smallest mixing angle $\theta_{13}$, 
the smallest neutrino mass $m_1$, and the Majorana phase $\alpha_{21}$. 
This statement is based on a Monte-Carlo study:
Treating unspecified $\mathcal{O}(1)$ parameters of the considered Froggatt-Nielsen
model as random variables, the observables of interest are sharply peaked
around certain central values.

We expect that these results hold beyond Froggatt-Nielsen flavour models.
An obvious example are extradimensional models which lead to the same type
of light neutrino mass matrix (cf.~\cite{Asaka:2003iy}). On the other hand, quark-lepton
mass hierarchies and the presently known neutrino observables cannot determine
the remaining observables in a model-independent way. This is illustrated
by the fact that our present knowledge about quark and lepton masses and
mixings is still consistent with $\theta_{13}\simeq0$ as well as with an inverted neutrino mass hierarchy (cf.~\cite{theta13}). As a 
consequence, further measurements of neutrino parameters will be able to falsify
certain patterns of flavour mixing and thereby provide valuable guidance for
the theoretical origin of quark and lepton mass matrices.

\bigskip\bigskip \noindent \textbf{Acknowledgements}

The authors thank G.~Altarelli, F.~Br\"ummer, G.~Ross, D.~Wark, W.~Winter and T.~Yanagida for helpful discussions and comments. This
work has been supported by the German Science Foundation (DFG) within the Collaborative
Research Center 676 ``Particles, Strings and the Early Universe''.

\appendix

\section{Analytic derivation of the Majorana phase $\alpha_{21}$ \label{app_alpha1}}

The complex phases of the $\mathcal{O}(1)$ coefficients in
the neutrino mass matrix $m_\nu$ and the lepton mass matrix $m_e$
are randomly distributed.
One would thus naively expect that also the Majorana phases
$\alpha_{21}$ and $\alpha_{31}$ in the PMNS matrix can take arbitrary values.
By contrast, the distribution of values for $\alpha_{21}$ that we obtain
from our numerical Monte-Carlo study, cf. Fig.~\ref{fig_masses}, clearly features a
prominent peak at $\alpha_{21} = \pi$.
In this appendix we shall demonstrate by means of a simplified example
how the structure of the neutrino mass matrix $m_\nu$ may partly
fix the phases of the corresponding mixing matrix $U$.

Consider the following simplified Majorana mass matrix $m_\nu$ for the light
neutrinos,
\begin{align}
m_\nu = v \begin{pmatrix}
\eta^2 & \eta e^{i\varphi}  & \eta \\ \eta e^{i\varphi} & 1 & 1 \\ \eta & 1 & 1\end{pmatrix}
\,,\quad
v = \frac{v^2_{\textrm{EW}} }{\bar{v}_{B-L}}
\,,
\label{eq:toym}
\end{align}
where $\varphi$ is an arbitrary complex phase between $0$ and $2\pi$.
For simplicity, let us neglect any effects on the mixing matrix $U$
from the diagonalization of $m_e$.
That is, we define $U$ such that
$U^T m_\nu U = \textrm{diag}\left(m_i\right)$,
with $m_i^2$ denoting the eigenvalues of $m_\nu^\dagger m_\nu$,
\begin{align}
\frac{m_{1,2}^2}{v^2}= & \: \eta^2 \sin^2\left(\varphi/2\right)
\left[2\mp\eta \left(5+3\cos\left(\varphi\right)\right)^{1/2}\right] +
\mathcal{O}\left(\eta^4\right)\,,\\
\frac{m_3^2}{v^2}= & \: 4\left(1 + \eta^2\left[1-\sin^2\left(\varphi/2\right)\right]\right) +
\mathcal{O}\left(\eta^4\right)\nonumber
\,.
\end{align}
Notice that the first two mass eigenvalues are nearly degenerate.
This is a consequence of the particular hierarchy pattern of
the matrix $m_\nu$ which originally stems from the equal flavour
charges of the $\textbf{5}_2^*$ and $\textbf{5}_3^*$ multiplets.
The relative sign of the $\mathcal{O}\left(\eta^3\right)$ contributions
to $m_1^2$ and $m_2^2$ eventually shows up again in entries of $U$, for instance,
\begin{align}
U_{11,12} = \mp\frac{2\left(5+3\cos\left(\varphi\right)\right)^{1/2}}{3 + e^{i\varphi}}
\,\exp\left(-\frac{i}{2}\textrm{Arg}\left[\mp z\right]\right)
+ \mathcal{O}\left(\eta\right)
\,.
\end{align}
with $z = 1-\cos\left(\varphi\right)-2i\sin\left(\varphi\right).$
The phase $\alpha_{21} = 2 \left(\textrm{Arg}\left[U_{12}/U_{11}\right]
\:\textrm{mod}\:\pi\right)$ in the matrix $U$ represents the analog
of the Majorana phase $\alpha_{21}$ in the PMNS matrix, cf. Eq.~\eqref{pmns}.
According to our explicit results for $U_{11}$ and $U_{12}$ it is independent
of the arbitrary phase $\varphi$ to leading order in $\eta$,
\begin{align}
\alpha_{21} \simeq 2 \left(\textrm{Arg}\left[- \exp\left(-\frac{i}{2}
\textrm{Arg}\left[+z\right]
+\frac{i}{2} \textrm{Arg}
\left[-z\right]
\right)\right] \:\textrm{mod}\:\pi\right) = \pi
\,.
\end{align}
In a similar way we may determine the phase analogous to the Majorana phase $\alpha_{31}$.
However, due to the hierarchy between the mass eigenvalues $m_1$ and $m_3$, the first and third
column of the matrix $U$ differ significantly from each other, thus  leading to a phase that
depends on $\varphi$ at all orders of $\eta$.

Including corrections to all orders in $\eta$ and scanning over the phase $\varphi$ numerically shows
that the maximal possible deviation of $\alpha_{21}$ from $\pi$ is, in fact,
of order $\eta^4$.
Adding more complex phases to the matrix $m_\nu$ in Eq.~\eqref{eq:toym}
gradually smears out the peak in the distribution of $\alpha_{21}$ values.
The distribution that is reached in the case of six different phases
is already very similar to the one in Fig.~\ref{fig_masses}.
We conclude that despite the need for corrections
the rough picture sketched in this appendix remains valid:
The hierarchy pattern of the neutrino mass matrix directly
implies that $\alpha_{21}$ tends to be close to $\alpha_{21} = \pi$.


\begin{thebibliography}{99} 

\bibitem{Raby:2008gh}
For a review and references see, for example, \\
S.~Raby,
  %``SUSY GUT Model Building,''
  Eur.\ Phys.\ J.\  C {\bf 59}, 223 (2009)
  %[arXiv:0807.4921 [hep-ph]].
  %%CITATION = EPHJA,C59,223;%%

\bibitem{Froggatt:1978nt}
  C.~D.~Froggatt and H.~B.~Nielsen,
  %``Hierarchy of Quark Masses, Cabibbo Angles and CP Violation,''
  Nucl.\ Phys.\  B {\bf 147}, 277 (1979)
  %%CITATION = NUPHA,B147,277;%%

\bibitem{extradimensions}
%\bibitem{Grossman:1999ra}
  Y.~Grossman and M.~Neubert,
  %``Neutrino masses and mixings in nonfactorizable geometry,''
  Phys.\ Lett.\  B {\bf 474}, 361 (2000);
  %[arXiv:hep-ph/9912408].
  %%CITATION = PHLTA,B474,361;%%
%
%\bibitem{Gherghetta:2000qt}
  T.~Gherghetta and A.~Pomarol,
  %``Bulk fields and supersymmetry in a slice of AdS,''
  Nucl.\ Phys.\  B {\bf 586}, 141 (2000);
  %[arXiv:hep-ph/0003129].
  %%CITATION = NUPHA,B586,141;%%
%
%\bibitem{Buchmuller:2006ik}
  W.~Buchmuller, K.~Hamaguchi, O.~Lebedev and M.~Ratz,
  %``Supersymmetric Standard Model from the Heterotic String (II),''
  Nucl.\ Phys.\  B {\bf 785}, 149 (2007)
 % [arXiv:hep-th/0606187].
  %%CITATION = NUPHA,B785,149;%%

\bibitem{Froggatt-Nielsen+SU(5)}
%\bibitem{Sato:1997hv}
  J.~Sato and T.~Yanagida,
  %``Large lepton mixing in a coset space family unification on E(7) / SU(5) x
  %U(1)**3,''
  Phys.\ Lett.\  B {\bf 430}, 127 (1998);
  %[arXiv:hep-ph/9710516].
  %%CITATION = PHLTA,B430,127;%%
%
%\bibitem{Irges:1998ax}
  N.~Irges, S.~Lavignac and P.~Ramond,
  %``Predictions from an anomalous U(1) model of Yukawa hierarchies,''
  Phys.\ Rev.\  D {\bf 58}, 035003 (1998)
  %[arXiv:hep-ph/9802334].
  %%CITATION = PHRVA,D58,035003;%%

\bibitem{Buchmuller:1998zf}
  W.~Buchmuller and T.~Yanagida,
  %``Quark lepton mass hierarchies and the baryon asymmetry,''
  Phys.\ Lett.\  B {\bf 445}, 399 (1999)
  %[arXiv:hep-ph/9810308].
  %%CITATION = PHLTA,B445,399;%%

\bibitem{Vissani:1998xg}
  F.~Vissani,
  %``Large mixing, family structure, and dominant block in the neutrino mass
  %matrix,''
  JHEP {\bf 9811}, 025 (1998)
  %[arXiv:hep-ph/9810435].
  %%CITATION = JHEPA,9811,025;%%

\bibitem{random_coeff}
%\bibitem{Hall:1999sn}
  L.~J.~Hall, H.~Murayama and N.~Weiner,
  %``Neutrino mass anarchy,''
  Phys.\ Rev.\ Lett.\  {\bf 84}, 2572 (2000);
  %[arXiv:hep-ph/9911341].
  %%CITATION = PRLTA,84,2572;%%
%
%\bibitem{Sato:2000kj}
  J.~Sato and T.~Yanagida,
  %``Low-energy predictions of lopsided family charges,''
  Phys.\ Lett.\  B {\bf 493}, 356 (2000);
 % [arXiv:hep-ph/0009205].
  %%CITATION = PHLTA,B493,356;%%
%
%\bibitem{Vissani:2001im}
  F.~Vissani,
  %``Expected properties of massive neutrinos for mass matrices with a dominant
  %block and random coefficients order unity,''
  Phys.\ Lett.\  B {\bf 508}, 79 (2001)
  %[arXiv:hep-ph/0102236].
  %%CITATION = PHLTA,B508,79;%%
%

\bibitem{Abe:2011sj}
  K.~Abe {\it et al.}  [T2K Collaboration],
  %``Indication of Electron Neutrino Appearance from an Accelerator-produced
  %Off-axis Muon Neutrino Beam,''
  Phys.\ Rev.\ Lett.\  {\bf 107}, 041801 (2011)
  %[arXiv:1106.2822 [hep-ex]].
  %%CITATION = PRLTA,107,041801;%%

%\cite{Adamson:2011qu}
\bibitem{Adamson:2011qu}
  P.~Adamson {\it et al.} [MINOS Collaboration],
  %``Improved search for muon-neutrino to electron-neutrino oscillations in MINOS,''
  Phys.\ Rev.\ Lett.\  {\bf 107}, 181802 (2011).
  %[arXiv:1108.0015 [hep-ex]].

\bibitem{doublechooz}
Talk given by H.\ de Kerret at the Sixth International Workshop on Low Energy Neutrino Physics (LowNu11) at Seoul, Korea during November 9-12, 2011


\bibitem{Nakamura:2010zzi}
  K.~Nakamura {\it et al.}  [Particle Data Group],
  %``Review of particle physics,''
  J.\ Phys.\ G {\bf 37}, 075021 (2010)
  %%CITATION = JPHGB,G37,075021;%%

\bibitem{cosmo}
%\bibitem{Buchmuller:2010yy}
  W.~Buchmuller, K.~Schmitz and G.~Vertongen,
  %``Matter and Dark Matter from False Vacuum Decay,''
  Phys.\ Lett.\  B {\bf 693}, 421 (2010);
  %[arXiv:1008.2355 [hep-ph]].
  %%CITATION = PHLTA,B693,421;%%
%
%\bibitem{Buchmuller:2011mw}
  W.~Buchmuller, K.~Schmitz and G.~Vertongen,
  %``Entropy, Baryon Asymmetry and Dark Matter from Heavy Neutrino Decays,''
  Nucl.\ Phys.\  B {\bf 851}, 481 (2011);
  %[arXiv:1104.2750 [hep-ph]].
  %%CITATION = NUPHA,B851,481;%%
%
% No SPIRES record found for cite request prep:2011xx
%\cite{Buchmuller:2012wn}
%\bibitem{Buchmuller:2012wn} 
  W.~Buchmuller, V.~Domcke and K.~Schmitz,
  %``Spontaneous B-L Breaking as the Origin of the Hot Early Universe,''
  arXiv:1202.6679 [hep-ph].
  %%CITATION = ARXIV:1202.6679;%%


%\cite{Cowan:1998ji}
\bibitem{Cowan:1998ji}
 G.~Cowan,
 ``Statistical data analysis,''
%\href{http://www.slac.stanford.edu/spires/find/hep/www?irn=3988171}{SPIRES entry}
{  Oxford, UK: Clarendon (1998) 197 p} 

\bibitem{:2006uk}
 [Planck Collaboration],
 %G.~Efstathiou, C.~Lawrence, J.~Tauber {\it et al.}  [Planck Collaboration],
 %Planck~Science~Team [Planck Collaboration],
 %``The Scientific programme of planck,''
 ESA-SCI(2005)1 (2006),
 arXiv:astro-ph/0604069
 %%CITATION = ASTRO-PH/0604069;%% 

%\cite{Beck:2010zzb}
\bibitem{Beck:2010zzb}
 M.~Beck  [KATRIN Collaboration],
 %``The KATRIN Experiment,''
 J.\ Phys.\ Conf.\ Ser.\  {\bf 203} (2010) 012097
 %[arXiv:0910.4862 [nucl-ex]].
 %%CITATION = 00462,203,012097;%% 

%\cite{KlapdorKleingrothaus:2001ke}
\bibitem{KlapdorKleingrothaus:2001ke}
  H.~V.~Klapdor-Kleingrothaus, A.~Dietz, H.~L.~Harney, I.~V.~Krivosheina,
  %``Evidence for neutrinoless double beta decay,''
  Mod.\ Phys.\ Lett.\  {\bf A16}, 2409-2420 (2001)
  %[hep-ph/0201231].

%\cite{Meierhofer:2011zz}
\bibitem{Meierhofer:2011zz}
 G.~Meierhofer  [GERDA Collaboration],
 %``Gerda: A New Neutrinoless Double Beta Experiment Using Ge-76,''
 J.\ Phys.\ Conf.\ Ser.\  {\bf 312} (2011) 072011.
 %%CITATION = 00462,312,072011;%% 

%\cite{Fukugita:1986hr}
\bibitem{Fukugita:1986hr}
  M.~Fukugita, T.~Yanagida,
  %``Baryogenesis Without Grand Unification,''
  Phys.\ Lett.\  {\bf B174}, 45 (1986).


\bibitem{leptogenesis}
%\bibitem{Buchmuller:2005eh}
  For reviews containing the relevant formulae see, for example, \\
 W.~Buchmuller, R.~D.~Peccei and T.~Yanagida,
  %``Leptogenesis as the origin of matter,''
  Ann.\ Rev.\ Nucl.\ Part.\ Sci.\  {\bf 55}, 311 (2005);
 % [arXiv:hep-ph/0502169].
  %%CITATION = ARNUA,55,311;%%
%
%\bibitem{Davidson:2008bu}
  S.~Davidson, E.~Nardi and Y.~Nir,
  %``Leptogenesis,''
  Phys.\ Rept.\  {\bf 466}, 105 (2008)
  %[arXiv:0802.2962 [hep-ph]].
  %%CITATION = PRPLC,466,105;%%

\bibitem{epsilon}
%\bibitem{Davidson:2002qv}
  S.~Davidson and A.~Ibarra,
  %``A Lower bound on the right-handed neutrino mass from leptogenesis,''
  Phys.\ Lett.\  B {\bf 535}, 25 (2002);
  %[arXiv:hep-ph/0202239].
  %%CITATION = PHLTA,B535,25;%%
%
%\bibitem{Hamaguchi:2001gw}
  see also K.~Hamaguchi, H.~Murayama and T.~Yanagida,
  %``Leptogenesis from N dominated early universe,''
  Phys.\ Rev.\  D {\bf 65}, 043512 (2002)
  %[arXiv:hep-ph/0109030].
  %%CITATION = PHRVA,D65,043512;%%

%\cite{hep-ph/0210342}
\bibitem{masina}
%\bibitem{hep-ph/0210342} 
  G.~Altarelli, F.~Feruglio and I.~Masina,
  %``Models of neutrino masses: Anarchy versus hierarchy,''
  JHEP\ {\bf 0301}, 035  (2003);
  %[hep-ph/0210342].
  %%CITATION = JHEPA,0301,035;%%
%
%\cite{hep-ph/0501166}
%\bibitem{hep-ph/0501166} 
  I.~Masina and C.~A.~Savoy,
  %``On power and complementarity of the experimental constraints on seesaw models,''
  Phys.\ Rev.\ D\ {\bf 71}, 093003  (2005)
  [hep-ph/0501166].
  %%CITATION = PHRVA,D71,093003;%%

\bibitem{winter}
%\cite{arXiv:0707.2379}
%\bibitem{arXiv:0707.2379} 
  F.~Plentinger, G.~Seidl and W.~Winter,
  %``The Seesaw mechanism in quark-lepton complementarity,''
  Phys.\ Rev.\ D\ {\bf 76}, 113003  (2007);
  %[arXiv:0707.2379 [hep-ph]].
  %%CITATION = PHRVA,D76,113003;%%
%
%\cite{hep-ph/0612169}
%\bibitem{hep-ph/0612169} 
  F.~Plentinger, G.~Seidl and W.~Winter,
  %``Systematic parameter space search of extended quark-lepton complementarity,''
  Nucl.\ Phys.\ B\ {\bf 791}, 60  (2008)
  %[hep-ph/0612169].
  %%CITATION = NUPHA,B791,60;%%

%\cite{Asaka:2003iy}
\bibitem{Asaka:2003iy}
  T.~Asaka, W.~Buchmuller, L.~Covi,
  %``Quarks and leptons between branes and bulk,''
  Phys.\ Lett.\  {\bf B563}, 209-216 (2003)
  %[hep-ph/0304142].

\bibitem{theta13}
%\cite{Altarelli:2010gt}
For a review and references see, for example,\\ 
%\bibitem{Altarelli:2010gt}
  G.~Altarelli, F.~Feruglio,
  %``Discrete Flavor Symmetries and Models of Neutrino Mixing,''
  Rev.\ Mod.\ Phys.\  {\bf 82}, 2701-2729 (2010);
  %[arXiv:1002.0211 [hep-ph]].
%
%\cite{Ishimori:2010au}
%\bibitem{Ishimori:2010au}
  H.~Ishimori, T.~Kobayashi, H.~Ohki, Y.~Shimizu, H.~Okada, M.~Tanimoto,
  %``Non-Abelian Discrete Symmetries in Particle Physics,''
  Prog.\ Theor.\ Phys.\ Suppl.\  {\bf 183}, 1-163 (2010)
  %[arXiv:1003.3552 [hep-th]].


\end{thebibliography}
\end{document}